\documentclass[twocolumn,showpacs,amsmath,amssymb]{revtex4-2}

\usepackage{amsmath}
\usepackage{graphicx}
\usepackage{amsfonts}
\usepackage{dcolumn}
\usepackage{bm}
\usepackage{color}

\begin{document}

\title{Activity-induced Nonequilibrium Vaporization Leads to Reentrant Phase Separation}

\author{Jie Su,$^{*}$ Mengkai Feng}
\thanks{These authors have contributed equally to this work.}
\author{Huijun Jiang}
\thanks{E-mail: hjjiang3@ustc.edu.cn}
\author{Zhonghuai Hou}
\thanks{E-mail: hzhlj@ustc.edu.cn}
\affiliation{Department of Chemical Physics \& Hefei National Laboratory for Physical Sciences at Microscales, University of Science and Technology of China, Hefei, Anhui 230026, China}
\date{\today}

\begin{abstract}
Active Brownian particles (ABPs) with pure repulsion is an ideal model to understand the effect of nonequilibrium on collective behaviors. It has long been established that activity can create effective attractions leading to motility-induced phase separation (MIPS), whose role is similar to that of (inverse) temperature in the simplest equilibrium system with attractive inter-particle interactions. Here, our theoretical analysis based on a kinetic theory of MIPS shows that a new type of activity-induced nonequilibrium vaporization is able to hinder the formation of dense phase when activity is large enough. Such nonequilibrium vaporization along with the activity-induced effective attraction thus lead to a MIPS reentrance. Numerical simulations verify such nonequilibrium effect induced solely by activity on phase behaviors of ABPs, and further demonstrate the dependence of MIPS on activity and the strength of inter-particle interaction predicted by our theoretical analysis. Our findings highlight the unique role played by the nonequilibrium nature of activity on phase behaviors of active systems, which may inspire deep insights into the essential difference between equilibrium and nonequilibrium systems.

\end{abstract}

\maketitle
Active systems consisting of self-propelled units have been widely discovered in nature on many scales, ranging from mesoscopic biological or manmade swimmers such as \emph{E. coli} and Janus spheres to macroscopic objects like fish, birds and horses\cite{a1,a2}. Due to the ability to take in and dissipate energy to drive themselves far from equilibrium\cite{a3}, active systems provide ideal model systems to investigate the effect of nonequilibrium on collective behaviors beyond their equilibrium counterparts\cite{a4,a5,a6,a7,a8,a9,a10,a11,a12,a13,a14,a15}. As one of the simplest active systems, active Brownian particles (ABPs) with pure repulsion without any attraction can spontaneously undergo phase separation between dense and dilute fluid phases\cite{MIPS1,MIPS2}. Such phase separation resulting solely from the intrinsically nonequilibrium property, i.e., activity, is so-called motility-induced phase separation (MIPS)\cite{MIPS1}.

Generally, for ABPs with pure repulsion, they tend to accumulate where they move more slowly and will slow down at high density for steric reasons, which then creates effective attractions leading to MIPS\cite{MIPS2}. Quite interestingly, it has been found that the phase diagram of ABPs\cite{MIPS2,MIPS3,MIPS5,MIPS12} is nearly the same as that of the simplest equilibrium system with attractive inter-particle interactions\cite{eqPS1,eqPS2,eqPS3,eqPS4}, except that the role of (inverse) temperature is replaced by activity. Is that all about the effect of activity on the phase separation of ABPs? Will the nonequilibrium nature of activity bring unique phase behaviors to the simplest ABPs beyond providing an alternative origin for phase separation?

In this work, we report that activity can lead to a type of nonequilibrium vaporization which lacks an analogue in equilibrium systems. Our theoretical analysis based on a kinetic theory of MIPS in the simplest ABP system with purely repulsive interactions shows that, such activity-induced nonequilibrium vaporization brings a unique phase behavior to ABPs. That is, while activity-induced effective attraction leads to MIPS, activity-induced nonequilibrium vaporization hinders the formation of dense phase and thus results in a MIPS reentrance. Besides, we find that the ``softer" the repulsive interaction potential is, the ``stronger" the reentrance becomes. The MIPS reentrance is then verified by numerical simulations. Furthermore, the simulated binodal agrees well with the phase boundary derived by our theoretical analysis, and the theoretically predicted dependence of MIPS on the strength of inter-particle interactions is also demonstrated. 

\emph{Theoretical analysis}.-- We start from a minimal active fluid theory characterizing MIPS from a kinetics approach\cite{MIPS3,MIPS4,MIPS5}. The theory describes the steady state of phase separation with a dense clustering phase which is set as close packed and a dilute phase as homogeneous and isotropic. Particles transport between two phases through condensation and evaporation events with rates $k_{\rm in}$ and $k_{\rm out}$, respectively. Absorption rate $k_{\rm in}$ is assumed to be proportion to the dilute phase number density $\rho_g$ and the characteristic self-propulsion velocity $v_0$. For the vaporization, the theory sets that active particles immediately escape from the dense phase when the direction of the particle on the interface moves towards to dilute phase. Therefore the escape rate $k_{\rm out}$ is proportion to the rotational diffusion constants $D_r$.  Then the steady state condition is achieved through equating $k_{\rm in}$ and $k_{\rm out}$, leading to the dilute phase density $\rho_g = \frac{\pi \kappa^\prime D_r}{\sigma v_0}$ where $\kappa^\prime$ is a fitting parameter and $\sigma$ is the particle diameter. For two dimensional ABP system with purely repulsive inter-particle potential \cite{MIPS3, MIPS5} and Lennard-Jones potential \cite{MIPS4}, solutions of $\rho_g$ show very good agreement with simulated MIPS.
However, it should be noted that the particle current from cluster to dilute phase comprises not only vaporization events due to rotational-diffusion-dependent active motion, but also those due to active-motion-dependent translational diffusion. Good agreement between solutions of $\rho_g$ and simulated MIPS indicates the former one makes a major contribution to $k_{\rm out}$ for activity ranging around the transition point of MIPS (i.e., small and moderate activity). Yet for large activity, the contribution of translational diffusion process becomes progressively more important as will be shown by the following analysis.

To take account of the vaporization events of active-motion-dependent translational diffusion, we need to firstly obtain the diffusivity of particles in dense phase near the interface. In the gas phase, the translational diffusion coefficient is close to the effective one for a free active particle $D_0 = D_t + \frac{v_0^2}{2D_r}$, where $D_t$ is purely thermal diffusion constant \cite{MIPS2}. In dense phase the local diffusion coefficient $D[\rho] = D_t + \frac{v[\rho]^2 }{2D_r}$ is much smaller than $D_0$, where $[\rho]$ denotes an arbitrary functional of density field $\rho(\vec{r})$ and $v[\rho]$ is a phenomenological motility parameter satisfying $0<v[\rho]<v_0$ \cite{MIPS2}. Around the dilute-dense phase interface, it is convenient to introduce a parameter $\lambda$ to express the local diffusion as $D[\rho] = D_t + \frac{\lambda v_0^2}{D_r}$, wherein $\lambda \in (0,\frac{1}{2})$.
We do notice that the divide of evaporation event, translational and rotational diffusion, is a little ambiguous and there might be double counting in this context. At least for a single event that a particle leaves the dense phase, it is difficult to determine which type of diffusion plays a major role. Nevertheless, we use a fitting parameter $\kappa$ to compromise this defect, then the total evaporation current is written as (see details in the supplemental information, SI)
\begin{align}
j_{\rm out} =& D[\rho] \nabla\rho + \kappa \frac{D_r}{\sigma} \label{jout}\nonumber \\
 = & \left( D_t + \frac{\lambda v_0^2}{D_r} \right) \frac{\rho_d - \rho_g}{\sigma} + \kappa \frac{D_r}{\sigma},
\end{align}
where $\rho_d$ is the number density of the dense phase. Analogous to the procedure in Ref.\cite{MIPS4}, by integrating the angles of ABPs over the direction of self-propulsion toward the interface, the condensation current is written as $j_{\rm in} = {\rho_g v_0} / {\pi} $.

Based on the steady-state assumption, i.e., $j_{\rm in}=j_{\rm out}$, two transition points can be immediately derived as
\begin{align}
v^{\pm}_c =& \frac{D_r\sigma}{2\lambda(\rho_d - \rho_g)} \Bigg[ \frac{\rho_g}{\pi} \nonumber \\
& \pm \sqrt{\frac{\rho_g^2}{\pi^2} - 4 \frac{\lambda (\rho_d - \rho_g)}{D_r \sigma^2} \left[ D_t(\rho_d-\rho_g) + \kappa D_r \right] } \Bigg].
\end{align}
We recall that, in the previous understanding of MIPS taking only rotational-diffusion-dependent vaporization into account, there is only one transition point $v_c$, since $j_{\rm out}$ is independent on $v_0$ and $j_{\rm in}$ increases linearly with $v_0$ \cite{MIPS3}. Consequently, MIPS will be observed as $v_0>v_c$, and activity is considered to mainly provide a nonequilibrium origin for phase separation. Quite interestingly, Eq.(2) points out that, increasing activity not only leads to MIPS when $v_0$ passes $v_c^-$ but also destroys it when $v_0 > v_c^+$. Such reentrant MIPS of the simplest ABPs with purely repulsive inter-particle interactions indicates that, beyond the activity-induced effective attraction when only rotational-diffusion-dependent vaporization is considered, extra nonequilibrium vaporization events due to active-motion-dependent translational diffusion further contributes a contrary effect on phase separation by accelerating melting. Such unique phase behavior induced by the nonequilibrium nature of activity has never been observed in the equilibrium analogues with attractive inter-particle interactions.

There are several further predictions can be made about the interesting phase behavior. Firstly, the steady state assumption $j_{\rm in}=j_{\rm out}$ gives a quantitative measurement of the dilute phase density, $\rho_g = \left( \rho_d D[\rho]+\kappa D_r \right) / \left( D[\rho] + v_0\sigma / \pi \right)$.
Unlike the previous theory where dilute phase density decreases with $v_0$ non-monotonically, $\rho_g$ eventually increases to $\rho_d$ for very large activity. This result strongly implies a homogeneous state when activity is sufficiently large, which confirms the reentrance of MIPS from another point of view.

Secondly, the parameter $\lambda$ is physically dependent on the repulsive inter-particle potential $U(r) = \epsilon \hat{U}(r)$, where $\epsilon$ is the strength of the potential and $\hat{U}(r)$ is an arbitrary dimensionless function of particle distance. Larger $\epsilon$ brings stronger repulsive force which hinders particle motion in the dense phase, and consequently leads to smaller diffusivity and smaller $\lambda$. Based on this picture, we introduce a so-called ``instant diffusion coefficient'' \cite{17SM_Feng}, defined as $D_i = D_t + \frac{v_0^2/(2D_r)}{1 +(2\gamma D_r)^{-1} \sum_{j\neq i}\nabla_i^2 u_{ij} }$, to estimate the local diffusivity $D[\rho]$ and then $\lambda$. Comparing with the formula of $D[\rho]$, we notice that $\lambda$ should be inversely proportional to $a+\epsilon$, where $a$ is introduced as an interaction and activity independent parameter.
According to Eq.(2), $v_c^+$ prominently increases with interacting strength $\epsilon$, while on the contrary $v_c^-$ slowly decreases. As a result, MIPS would also be reentrant for finite interacting strength $\epsilon$, while  increasing $\epsilon$ would greatly broaden the parameter region for emergence of MIPS. It is thus another testable prediction of our theory to investigate the dependence of MIPS behavior on the interacting strength.

\begin{figure}
\centering
\includegraphics[width=0.75\columnwidth]{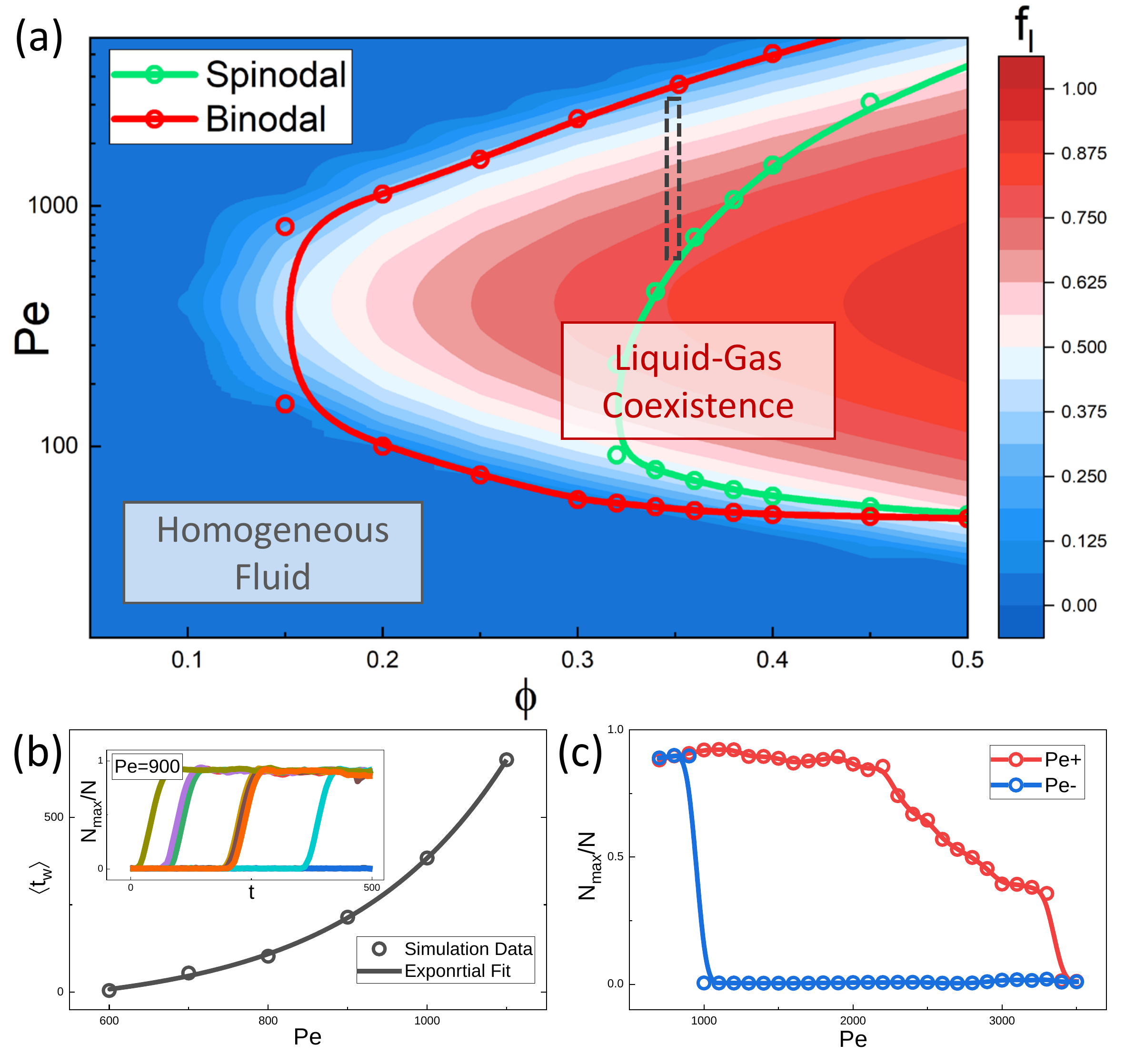}
\caption{(a) Phase diagram in the ${\rm Pe}-\phi$ plane. The colored background is obtained by the kinetics model with $\lambda=2.8\times10^{-4}$ and $\kappa=0.9$. $f_l$ is the fraction of dense phase. The red and green lines respectively represent the binodal and spinodal curves from simulations. (b) Dependence of the ensemble-averaged waiting time $\langle t_w\rangle$ for the nucleation on ${\rm Pe}$ for active systems at $\phi=0.35$ inside the gray dashed box in (a). Curves are the exponential fitting for the simulation data. The inset is the dependence of $N_{\rm max}/N$ with $N_{\rm max}$ the particle number of the largest cluster on $t$ for ten independent ensembles at $\phi=0.35$ and ${\rm Pe}=900$. (c) Hysteresis at $\phi=0.35$ including the gray dashed box in (a). Simulation data are reached from two initial conditions: the final configurations of systems at slightly smaller ${\rm Pe}$ (red line and symbols) and the ones at slightly larger ${\rm Pe}$ (blue line and symbols). }
\label{fig:result1}
\end{figure}

\emph{Numerical simulation}.-- In order to verify the unique phase behavior predicted by the aforementioned theoretical analysis, we perform simulations of a quasi two-dimensional system with size $L_x\times L_y$ and periodic boundary conditions consisting of $N$ spherical ABPs with diameter $\sigma$ and friction coefficient $\gamma$. The pairwise inter-particle potential is set as Weeks-Chandler-Andersen (WCA) potential $U({r}_{ij})=4\epsilon[(\sigma/r_{ij})^{12}-(\sigma/r_{ij})^6+1/4]$ for $r_{ij}<2^{1/6}\sigma$ and $U({r}_{ij})=0$ otherwise.
We set the translational diffusion coefficient $D_t=k_BT/\gamma$ so that the system satisfies the fluctuation-dissipation relation with $k_B$ the Boltzmann constant and $T$ the temperature. The rotational diffusion coefficient $D_r$ is coupled with the translational diffusivity as $D_r = 3D_t / \sigma^2$. In this work, we use two dimensionless variables as control parameters, i.e. the P\'{e}clet number ${\rm Pe}=v_0\sigma/D_t$ \cite{MIPS3} to characterize the activity of ABPs and the volume fraction $\phi=\pi\rho_0/(4\sigma^2)$ to describe the density of ABPs with $\rho_0=N/(L_x\times L_y)$ the averaged number density of the system (see details of the model in the SI).

The phase diagram in the ${\rm Pe}-\phi$ plane with $\epsilon=1$ is presented in Fig.~\ref{fig:result1}(a), wherein the color bar denotes the theoretical prediction of particle number fraction in dense phase $f_l = \frac{\rho_d}{\rho_0}  \frac{\rho_0 - \rho_g }{\rho_d - \rho_g}$ (since the number density of dense phase $\rho_d$ is not sensitive to the P\'{e}clet number and total density $\rho_0$, we assume that $\rho_d$ is constant and therefore, $f_l$ only depends on $\rho_g$ and $\rho_0$). It can be found that the simulated binodal (red dotted line) confirms with the phase boundary of the colored background predicted by our theoretical analysis. Likewise, theoretically predicted reentrance of MIPS can be observed as activity increases. For appropriate $\phi$ such as $\phi=0.35$, the active system changes from a single phase to a coexisting phase and reenters to a single phase as ${\rm Pe}$ increases (typical snapshots can be found in the SI). Besides of the binodal, the spinodal (green dotted line) can also be obtained by simulations too. It can be found that both the binodal and spinodal curves shift to lower $\phi$ first, but reenter to higher $\phi$ again after an inflexion, forming a brand new metastable state located in a much wider region at large ${\rm Pe}$.

\begin{figure}
\centering
\includegraphics[width=0.75\columnwidth]{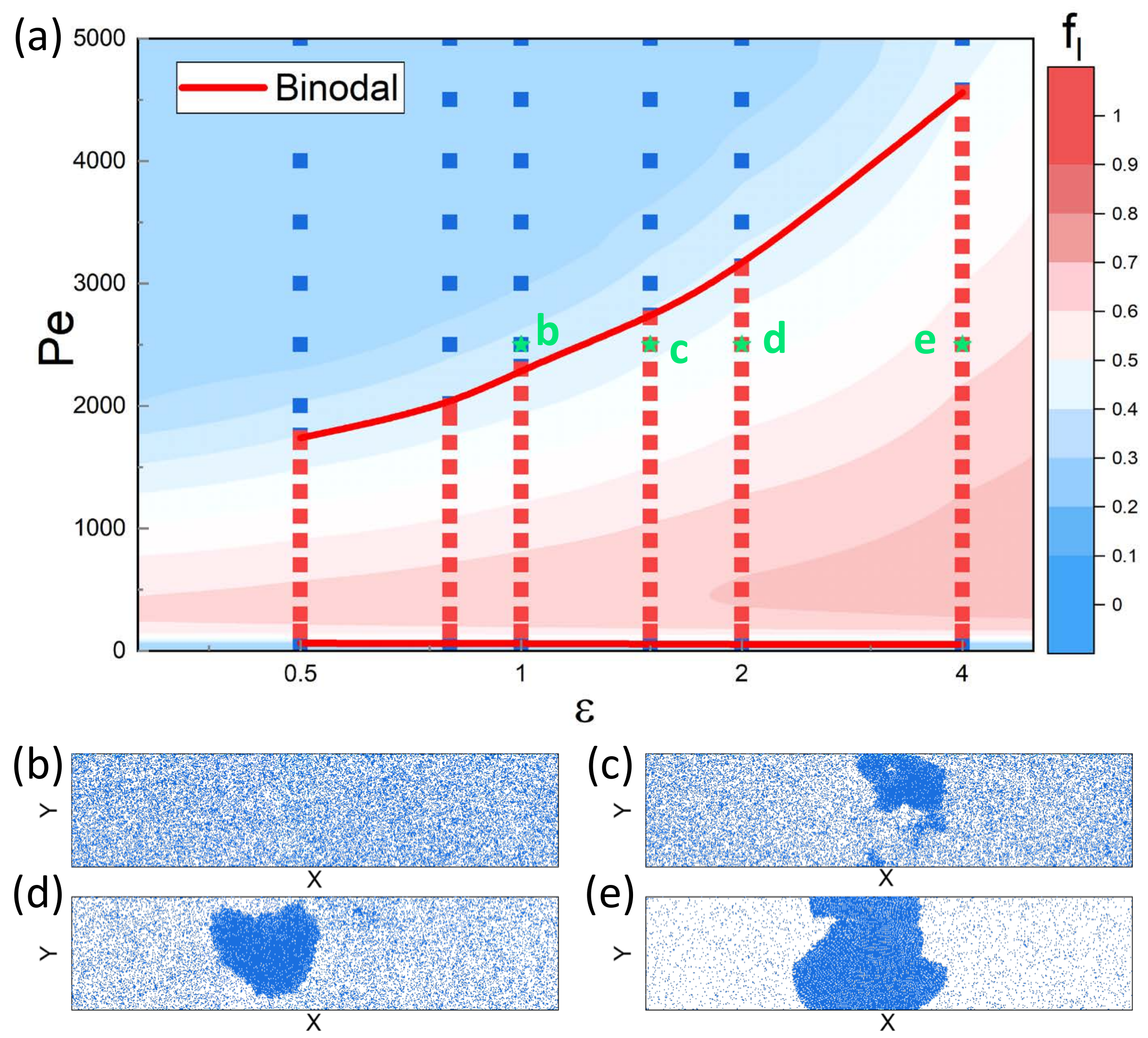}
\caption{Influence of the interaction strength. (a) Phase diagram in the ${\rm Pe}-\epsilon$ plane at $\phi=0.3$. The colored background is obtained by the kinetics model. The blue symbols represent the single phase, the red ones denote MIPS and the red-line boundaries is the binodal curves, which are all obtained by simulations. The steady-state configurations at ${\rm Pe}=2500$ and $\epsilon=1.0$ (b), $1.5$ (c), $2.0$ (d) and $4.0$ (e) are respectively marked as green stars in (a).}
\label{fig:result2}
\end{figure}

To investigate the nature of MIPS near the MIPS reentrance region, the growth process of the largest cluster (Fig.~\ref{fig:result1}(b)) and a hysteresis loop (Fig.~\ref{fig:result1}(c)) are focused. The inset of Fig.~\ref{fig:result1}(b) plots time series of $N_{\rm max}/N$ with $N_{\rm max}$ the particle number of the largest cluster for ten independent simulations with $\phi=0.35$ and ${\rm Pe}=900$ (time series for other parameters can be found in the SI). It can be observed that $N_{\rm max}/N$ increases from about 0 quickly after a waiting time $t_w$, indicating a first-order phase transition via nucleation \cite{MIPS3}. Noticing that the waiting time can be a good parameter to measure the nucleation barrier, the obtained ensemble-averaged waiting time $\langle t_w\rangle$ as a function of ${\rm Pe}$ is then presented in Fig.~\ref{fig:result1}(b). As ${\rm Pe}$ increases across the upper spinodal, $\langle t_w\rangle$ increases exponentially, demonstrating that systems with larger ${\rm Pe}$ above the upper spinodal must go over a higher nucleation barrier to attain MIPS. Besides, a very large hysteresis loop with $\phi=0.35$ can be found around the upper MIPS transition point as shown in Fig.~\ref{fig:result1}(c), demonstrating again the nucleation behavior of a discontinuous transition from a homogeneous initial state to MIPS\cite{MIPS6}. In short, the simulation results are consistent with our theoretical predictions, verifying that activity can bring unique phase behavior to the simplest ABP system.

Fig.~\ref{fig:result2}(a) verifies the theoretical prediction about the effect of interaction strength $\epsilon$ on MIPS with fixed $\phi=0.3$, where blue and red symbols indicate the single phase and the coexisting phase of MIPS, respectively. Phase boundaries between these two phases are the binodal (red lines), and the colored background is $f_l$ obtained by our theoretical analysis. Clearly, the upper binodal increases remarkably as $\epsilon$ increases, while the lower binodal stays nearly unchanged. Such observation agrees with the theoretical prediction very well. To take a close look at the influence of $\epsilon$, steady-state configurations of systems at ${\rm Pe}=2500$ and $\epsilon=1.0$, $1.5$, $2.0$ and $4.0$ are shown in Fig.~\ref{fig:result2}(b)-(e), respectively. As $\epsilon$ increases, the dense phase emerges with a relatively small cluster, and then grows to be a large cluster, indicating more and more pronounced MIPS. In other words, ``harder" repulsive interactions between ABPs will lead to ``weaker" reentrant MIPS while the ``softer" ones will result in ``stronger" reentrant MIPS, which further demonstrates the unique phase behavior induced by the nonequilibrium nature of activity.

In summary, reentrant MIPS induced solely by activity has been revealed. We showed both theoretically and numerically that such reentrant MIPS results from the competition between activity-induced effective attraction of ABPs preferring particle accumulation and activity-induced nonequilibrium vaporization hindering formation of large clusters. Our findings highlight the unique role played by the nonequilibrium nature of activity on phase behaviors of active systems, which may inspire deep insights into the essential difference between equilibrium and nonequilibrium systems.

This work is supported by MOST(2018YFA0208702), Innovation Program for Quantum Science and Technology (2021ZD0303306), NSFC (32090044, 21973085, 21833007, 21790350), Anhui Initiative in Quantum Information Technologies (AHY090200), and the Fundamental Research Funds for the Central Universities (WK2340000104).


\end{document}